\documentclass[journal]{IEEEtran}
\IEEEoverridecommandlockouts
\usepackage{cite}
\usepackage{amsmath,amssymb,amsfonts}
\usepackage{algorithm, algorithmic}

\usepackage[bookmarks=false]{hyperref}

\usepackage{amsthm}

\newtheorem{theorem}{Theorem}

\usepackage{graphicx}
\usepackage{textcomp}
\usepackage{xcolor}
\usepackage{multirow}
\usepackage{hyperref} 
\usepackage{adjustbox}
\usepackage{tabularx}
\usepackage{siunitx}
\usepackage{array}
\usepackage[caption=false,font=normalsize,labelfont=sf,textfont=sf]{subfig}
\usepackage{stfloats}
\usepackage{url}
\usepackage{verbatim}

\def\BibTeX{{\rm B\kern-.05em{\sc i\kern-.025em b}\kern-.08em
    T\kern-.1667em\lower.7ex\hbox{E}\kern-.125emX}}

\begin{document}

\title{PRZK-Bind: A Physically Rooted Zero-Knowledge Authentication Protocol for Secure Digital Twin Binding in Smart Cities
}

\author{
\IEEEauthorblockN{
Yagmur Yigit\IEEEauthorrefmark{1},
Mehmet Ali Erturk\IEEEauthorrefmark{1},
Kerem Gursu\IEEEauthorrefmark{2}, and 
Berk Canberk\IEEEauthorrefmark{1}}
\IEEEauthorblockA{\\
\IEEEauthorrefmark{1}School of Computing, Engineering and The Build Environment, Edinburgh Napier University, United Kingdom \\
 \IEEEauthorrefmark{2} BTS Group, Turkey 
\\
Email: \{yagmur.yigit, m.erturk, b.canberk\}@napier.ac.uk, kerem.gursu@btsgrp.com}
}

\markboth{Accepted by 2025 IEEE Global Communications Conference (GLOBECOM), ©2025 IEEE}%
{Shell \MakeLowercase{\textit{et al.}}}

\maketitle

\begin{abstract}
Digital twin (DT) technology is rapidly becoming essential for smart city ecosystems, enabling real-time synchronisation and autonomous decision-making across physical and digital domains. However, as DTs take active roles in control loops, securely binding them to their physical counterparts in dynamic and adversarial environments remains a significant challenge. Existing authentication solutions either rely on static trust models, require centralised authorities, or fail to provide live and verifiable physical-digital binding, making them unsuitable for latency-sensitive and distributed deployments. To address this gap, we introduce PRZK-Bind, a lightweight and decentralised authentication protocol that combines Schnorr-based zero-knowledge proofs with elliptic curve cryptography to establish secure, real-time correspondence between physical entities and DTs without relying on pre-shared secrets. Simulation results show that PRZK-Bind significantly improves performance, offering up to 4.5 times lower latency and 4 times reduced energy consumption compared to cryptography-heavy baselines, while maintaining false acceptance rates more than 10 times lower. These findings highlight its suitability for future smart city deployments requiring efficient, resilient, and trustworthy DT authentication.
\end{abstract}


\section{Introduction}
\label{sec:intro}
The rise of next-generation smart infrastructures, from autonomous transportation systems to intelligent energy grids, brings unprecedented secure and reliable communication requirements \cite{Digi24, BC1}. These systems, often distributed across wide urban areas, are characterised by strict latency, privacy, and resilience demands \cite{twinPort, BC4}. In particular, Digital Twin (DT) technology is rapidly emerging as a key enabler for real-time synchronisation and decision-making, offering virtual representations of physical entities to support autonomous operations \cite{YY22, Rev21}. As DTs increasingly participate in operational control loops, from traffic management to smart healthcare, secure and timely authentication between physical entities and their twins becomes a fundamental necessity \cite{SecSurvey22, IoT-YY}.
%

Recent efforts have explored various DT authentication and binding schemes to tackle this challenge. Attribute-based and pairing-based methods~\cite{RSAKA25} provide flexible access control but suffer from high computational costs, rendering them impractical for resource-constrained edge scenarios. Similarly, lightweight
%
designs such as PUF-based schemes~\cite{VeDTAut21} lower processing needs but struggle with stability in dynamic conditions. Furthermore, several works~\cite{GC23, GC24, BlockDT24} leverage DT-enhanced mechanisms in vehicular and IoT networks, yet they either rely on pre-shared symmetric secrets, centralised verification authorities, or do not ensure that the DT corresponds to a live physical entity at the time of interaction.
%
%
This gap is critical in smart city scenarios requiring real-time response and cross-domain collaboration. Without decentralised live binding, adversaries can impersonate DTs and disrupt operations. Moreover, the lack of lightweight, robust binding degrades key Quality of Service (QoS), including authentication latency, false acceptance rate (FAR), computation overhead, and energy consumption, essential for secure and fast communication in smart city systems \cite{Rev22, BC3, BC2}.

To address these challenges, we present PRZK-Bind, a novel protocol for secure physical–digital binding at the network edge. Unlike prior work, it needs no pre-shared secrets or central verifiers, ensuring real-time correspondence. PRZK-Bind combines a Schnorr-based zero-knowledge proof with lightweight elliptic curve operations to provide strong security and low latency, while resisting impersonation, replay, and key compromise attacks in adversarial settings.
Our main contributions are as follows:
\begin{itemize}
    \item We address the limits of existing DT authentication, which often rely on pre-shared secrets, central authorities, or lack live physical–digital checks.  
    
    \item We propose PRZK-Bind, a decentralised zero-knowledge protocol enabling lightweight mutual binding without a central authority.  
    
    \item We give a formal security analysis showing PRZK-Bind resists impersonation, replay, and key compromise, while ensuring forward secrecy.    
\end{itemize}

The rest of the paper is organised as follows. Section~\ref{sec:related} reviews the related literature. Section~\ref{sec:system} introduces the system model, and Section~\ref{sec:proposed} details the proposed authentication mechanism. Section~\ref{sec:security} presents the security analysis, while Section~\ref{sec:performance} discusses performance evaluation. Finally, Section~\ref{sec:conclusion} concludes the paper.

\section{Related Work}
\label{sec:related}

Many existing DT authentication schemes rely on static trust, pre-shared secrets, or central authorities, which prevents them from ensuring dynamic and real-time binding in distributed smart city environments.
Existing research falls into three categories. Cryptography-intensive methods~\cite{RSAKA25} offer strong security via attribute-based encryption and pairings but suffer from high computational and latency overhead, making them unsuitable for edge use. Physical identity-based schemes~\cite{VeDTAut21, SchnorrIoT24} reduce computational cost by leveraging hardware-derived identities but do not ensure live binding. Decentralised solutions~\cite{BlockDT24, GC22, GC24} improve scalability and trust management through blockchain and distributed signatures but often overlook real-time verification between physical objects and their digital twins. Edge-oriented frameworks such as~\cite{Rev3} provide lightweight and modular DT architectures but do not address secure physical–digital binding.
A comparison of these works and their limitations is summarised in Table~\ref{tab:comparison}. As clearly summarised, existing approaches either impose heavy cryptographic overhead, rely on centralised components, or fail to ensure that the digital twin truly reflects a live physical entity during interaction. In contrast, our PRZK-Bind protocol is designed to overcome these limitations by providing decentralised, lightweight, and live binding authentication tailored for smart city edge networks.

\begin{table}[t]
\centering
\caption{Comparison of Existing Solutions}
\label{tab:comparison}
\begin{tabular}{l c c c c}
    \hline \hline
    Work & Decentralised & Lightweight & Live Binding & Edge-Deployable \\
    \hline \hline
    \cite{RSAKA25} & $\times$ & $\times$ & $\times$ & $\times$ \\
    \cite{VeDTAut21} & $\checkmark$ & $\checkmark$ & $\times$ & $\checkmark$ \\
    \cite{GC23} & $\times$ & $\checkmark$ & $\times$ & $\checkmark$ \\
    \cite{BlockDT24} & $\checkmark$ & $\times$ & $\times$ & $\times$ \\
    \cite{SchnorrIoT24} & $\checkmark$ & $\checkmark$ & $\times$ & $\checkmark$ \\
    \cite{GC22} & $\checkmark$ & $\checkmark$ & $\times$ & $\checkmark$ \\
    \cite{Rev3} &  $\checkmark$ & $\checkmark$ & $\times$ & $\checkmark$ \\
    Ours & $\checkmark$ & $\checkmark$ & $\checkmark$ & $\checkmark$ \\
    \hline \hline
\end{tabular}
\end{table}

\section{System Model}
\label{sec:system}
We consider a smart city setting where DTs support real-time monitoring and autonomous control. As an example, we model a smart traffic light and its DT at the edge or a fog node. The goal is to establish secure identity binding before any synchronised operation or data exchange.
The proposed system consists of the following key components:
\begin{itemize}
    \item $\mathcal{P}$: A physical object, such as a smart traffic light, with sensors and a hardware-derived identity. 
    \item $\mathcal{D}$: The digital twin of $\mathcal{P}$, deployed at the edge to manage and emulate the state and behaviour of $\mathcal{P}$ in real-time.
    \item $\mathcal{CA}$: A credential authority used only at initialisation for parameters and identity provisioning. After registration, PRZK-Bind runs without it.  
    \item $\mathcal{E}$: The communication environment, assumed to be an untrusted and open wireless medium.   
\end{itemize}
%
We assume the following communication properties:
\begin{itemize}
    \item $\mathcal{P}$ and $\mathcal{D}$ communicate over an insecure wireless medium open to adversarial monitoring and manipulation.  
    \item $\mathcal{D}$ can act on behalf of $\mathcal{P}$ only after successful authentication.  
    \item Authentication and binding must be decentralised and locally verifiable, without real-time support from $\mathcal{CA}$.  
\end{itemize}
%
We define the following cryptographic components:
\begin{itemize}
    \item $\mathcal{S}_{p}$: A unique and unforgeable identifier of the physical object $\mathcal{P}$, derived from sources such as a Physical Unclonable Functions (PUFs) or RF-fingerprints.
    \item $pk_p$: The public key of $\mathcal{P}$ computed as $pk_p = g^{H_1(\mathcal{S}_p)}$ using a secure hash-based key derivation.
    \item $sk_d, pk_d$: The private-public key pair generated by $\mathcal{D}$ using elliptic curve cryptography (ECC).
    \item $H_1, H_2$: Collision-resistant hash functions modelled as random oracles.
    \item $g$: A generator of cyclic group $\mathbb{G}$ of prime order $q$, used for key establishment.
\end{itemize}
%
The system operates through four primary phases:
\subsubsection*{1) Initialization}
$\mathcal{CA}$ defines global system parameters $\{g, q, \mathbb{G}, H_1, H_2\}$ and securely distributes them to $\mathcal{P}$ and $\mathcal{D}$. The physical object $\mathcal{P}$ derives its public key as:
\begin{equation}
    pk_p = g^{H_1(\mathcal{S}_p)}
\end{equation}
Meanwhile, the digital twin $\mathcal{D}$ selects a secret $sk_d \in \mathbb{Z}_q^*$ and computes:
\begin{equation}
    pk_d = g^{sk_d}
\end{equation}

\subsubsection*{2) Registration}
During the registration phase, both $\mathcal{P}$ and $\mathcal{D}$ register their public identities to $\mathcal{CA}$, which creates a verifiable binding record $\zeta$ as:
\begin{equation}
    \zeta = H_2(pk_p \parallel pk_d \parallel T)
\end{equation}
where $T$ is a trusted timestamp that anchors the session to a verifiable temporal context and $\zeta$ serves as the reference token during authentication.

\subsubsection*{3) Authentication}
When a session begins, $\mathcal{D}$ proves knowledge of $sk_d$ through a zero-knowledge proof, and $\mathcal{P}$ reveals a hash of its physical identity. A session key is computed as:
\begin{equation}
    K_{pd} = H_1(pk_p^{sk_d} \parallel \zeta)
\end{equation}

\subsubsection*{4) Verification and Binding}
The binding phase completes only if both the Schnorr proof and physical identity check pass, ensuring $\mathcal{D}$ truly matches $\mathcal{P}$ before any operation.
We assume an active adversary $\mathcal{A}$ with the following capabilities:
\begin{itemize}
    \item Interception, modification, and replay of messages exchanged between $\mathcal{P}$ and $\mathcal{D}$.
    \item Impersonation attempts targeting either $\mathcal{P}$ or $\mathcal{D}$.
    \item Key compromise attempts targeting $sk_d$ or cloning attempts targeting $\mathcal{S}_p$.
\end{itemize}

The proposed protocol is designed to establish mutual trust, prevent identity forgery, and ensure cryptographic resilience under dynamic and distributed smart city conditions.

\section{Proposed Authentication Protocol}
\label{sec:proposed}
This section presents PRZK-Bind, an authentication protocol designed to establish a secure, lightweight, and verifiable identity binding between a physical object $\mathcal{P}$ and its digital twin $\mathcal{D}$ in a smart city environment. The mechanism incorporates physical-layer identity features, ECC, and a Schnorr-based zero-knowledge proof to ensure mutual authentication under latency and resource constraints.


We adopt the Schnorr Identification Protocol \cite{Schnorr89} as the zero-knowledge foundation for our scheme based on the following considerations:
\begin{itemize}
    \item The protocol requires only two modular exponentiations and one hash computation per session, significantly reducing computational overhead compared to zk-SNARKs or Bulletproofs.
    \item Schnorr is provably secure under the hardness of the Discrete Logarithm Problem (DLP) in the Random Oracle Model.
    \item It integrates naturally within ECC settings and can be rendered non-interactive via the Fiat–Shamir heuristic, supporting asynchronous deployments in edge-based digital twin systems.
\end{itemize}

Let $G$ denote a cyclic group of prime order $q$ with generator $g$. The physical entity $\mathcal{P}$ exposes a unique and unclonable identity $\mathcal{S}_p$, used to derive a hash-based public key:
\begin{equation}
    pk_p = g^{H_1(\mathcal{S}_p)} \in G
\label{eq:pkp}
\end{equation}

The digital twin $\mathcal{D}$ independently generates an ECC key pair $(sk_d, pk_d)$, where:
\begin{equation}
    pk_d = g^{sk_d}
\end{equation}

\begin{figure}[t]
    \centering
    \includegraphics[width=3in]{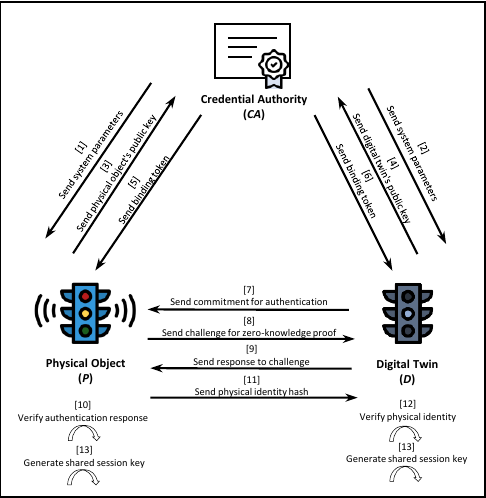}
    \caption{Overview of the decentralised mutual authentication and physical-digital binding process in PRZK-Bind.}
    \label{fig:overview}
\end{figure}

The operation of PRZK-Bind involves three core phases: registration, authentication, and binding. As illustrated in Fig.~\ref{fig:overview}, the protocol ensures that the physical object and the digital twin participate in a mutual authentication process without relying on centralised verification during live operations.

During the registration phase, both public keys are submitted to $\mathcal{CA}$, which computes a binding commitment $\zeta$:
\begin{equation}
    \zeta = H_2(pk_p \parallel pk_d \parallel T)
\label{eq:zeta}
\end{equation}
where $T$ denotes a trusted timestamp.

\subsubsection*{\textbf{Authentication Procedure}}
Once a session begins, $\mathcal{D}$ must prove possession of $sk_d$ using a Schnorr-style challenge–response mechanism:
\begin{itemize}
    \item $\mathcal{D}$ selects a random $r \in \mathbb{Z}_q$, computes $\alpha = g^r$, and transmits $\alpha$ to $\mathcal{P}$.
    \item $\mathcal{P}$ generates a challenge $c = H_1(\alpha \parallel \zeta)$.
    \item $\mathcal{D}$ computes the response $z = r + c \cdot sk_d \bmod q$.
    \item $\mathcal{P}$ verifies:
        \begin{equation}
            g^z \stackrel{?}{=} \alpha \cdot pk_d^c
        \label{eq:schnorr_verification}
        \end{equation}
\end{itemize}

\subsubsection*{\textbf{Physical Identity Verification}}
In addition, $\mathcal{P}$ transmits both $H_1(\mathcal{S}_p)$ for identity verification and $g^{r_p}$ as a fresh ephemeral contribution, ensuring that the binding process is tied to session-specific randomness in each execution:
\begin{equation}
    g^{H_1(\mathcal{S}_p)} \stackrel{?}{=} pk_p
\end{equation}
This two-sided verification ensures that both entities are authenticated and cryptographically bound.

\subsubsection*{\textbf{Session Key Derivation}}
If both validation steps succeed, the parties derive a shared session key:
\begin{equation}
    K_{pd} = H_1\big((pk_p \cdot g^{r_p})^{sk_d} \parallel \zeta\big)
\end{equation}
Here, $\mathcal{P}$ generates an ephemeral randomness $r_p \in \mathbb{Z}_q$ for each session and publishes $g^{r_p}$ alongside $H_1(\mathcal{S}_p)$. This ensures that the Diffie–Hellman cannot be computed by an eavesdropper, since deriving $(pk_p \cdot g^{r_p})^{sk_d}$ requires solving the 
Computational Diffie–Hellman (CDH) problem. The use of ephemeral randomness transforms the design into a static–ephemeral Diffie–Hellman variant, providing provable indistinguishability of the session key under the Decisional Diffie–Hellman (DDH) assumption.

\begin{figure}[t]
    \centering
    \includegraphics[width=2.3in]{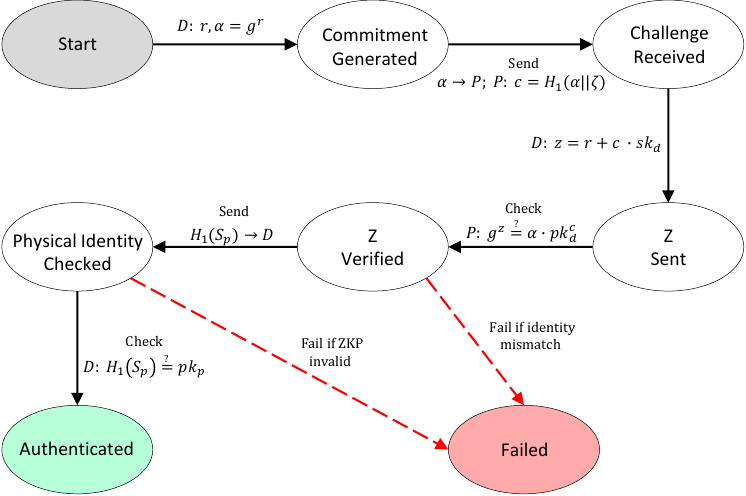}
    \caption{State-transition diagram of the proposed PRZK-Bind authentication protocol between a physical object and its digital twin.}
    \label{fig:state}
\end{figure}

To represent the sequential behaviour of the PRZK-Bind protocol, we illustrate its operation as a state-transition diagram in Fig.~\ref{fig:state}. Each state corresponds to a key step in the mutual authentication process between $\mathcal{P}$ and $\mathcal{D}$. Transitions are triggered by protocol-specific actions such as commitment generation, challenge creation, response computation, and verification. The validation of the Schnorr proof and the physical identity hash are modelled as critical decision points. If either check fails, the system transitions to a failure state. Upon successful execution, the final authentication state confirms the secure binding between $\mathcal{P}$ and its corresponding digital twin $\mathcal{D}$.
The proposed mechanism provides the following guarantees:
\begin{itemize}
    \item \textbf{Zero-Knowledge:} The Schnorr protocol ensures that no information about $sk_d$ or $\mathcal{S}_p$ is revealed during authentication.
    \item \textbf{Replay Protection:} The challenge is session-specific and tightly bound to $\zeta$ and $\alpha$, preventing reuse of previous proofs.
    \item \textbf{Impersonation Resistance:} An adversary lacking $sk_d$ cannot compute a valid response that satisfies Eq.~\ref{eq:schnorr_verification}.
    \item \textbf{Mutual Binding:} The protocol enforces bidirectional trust by requiring both $\mathcal{P}$ and $\mathcal{D}$ to actively participate in the authentication process. $\mathcal{P}$ proves its physical identity via $\mathcal{S}_p$, while $\mathcal{D}$ demonstrates knowledge of $sk_d$. This ensures that only a valid digital twin can be bound to the intended physical entity.
\end{itemize}
%
PRZK-Bind is well-suited for smart city infrastructures where autonomous devices need secure and efficient authentication. Its lightweight design and offline verifiability support use cases like traffic systems, sensor networks, and emergency response.

\section{Security Analysis}
\label{sec:security}
We analyse the security of PRZK-Bind under an active adversary $\mathcal{A}$ that can fully control the communication channel by intercepting, replaying, injecting, and modifying messages. The objective of $\mathcal{A}$ is to impersonate either party or to derive the session key without holding legitimate credentials.

\subsection{Security Definition}

We adopt the Real-or-Random (ROR) authenticated key-exchange model. In this setting, $\mathcal{A}$ observes all public transcripts ($\alpha, c, z, H_1(\mathcal{S}_p), g^{r_p}, pk_p, pk_d, \zeta$) and can issue test-session queries. The scheme is secure if $\mathcal{A}$ cannot distinguish the real session key $K_{pd}$ from random with non-negligible advantage. Security relies on the hardness of CDH/DDH in $G$ and the random oracle assumption for $H_1$ and $H_2$.

\subsection{Impersonation Resistance}
\begin{theorem}
An adversary $\mathcal{A}$ cannot impersonate the digital twin $\mathcal{D}$ to the physical entity $\mathcal{P}$ without knowledge of the private key $sk_d$, under the assumption that DLP is hard in $\mathbb{G}$.
\end{theorem}

\begin{proof}
To impersonate $\mathcal{D}$, the adversary must produce a valid response $z$ that satisfies the Schnorr verification equation:
\begin{equation}
    g^z \stackrel{?}{=} \alpha \cdot pk_d^c
\end{equation}
Given $\alpha = g^r$ and $c = H_1(\alpha \parallel \zeta)$, producing $z = r + c \cdot sk_d$ without knowledge of $sk_d$ requires solving the DLP, which is assumed to be infeasible in polynomial time. Therefore, the advantage of $\mathcal{A}$ in developing a valid response is negligible.
\end{proof}

\subsection{Replay Attack Resistance}
\begin{theorem}
The PRZK-Bind protocol is secure against replay attacks. A previously recorded transcript $(\alpha, c, z)$ cannot be reused in future sessions to impersonate $\mathcal{D}$.
\end{theorem}
\begin{proof}
Each session uses fresh randomness $r$ and a new challenge $c = H_1(\alpha \parallel \zeta)$. The temporary value $\alpha$ is derived from $r$ and is never reused. Since $H_1$ is modelled as a random oracle, $c$ will differ across sessions, even for the same entity pair. Replaying an old $(\alpha, c, z)$ tuple will fail because the new challenge will not match the old one, thus invalidating the proof.
\end{proof}

\subsection{Mutual Binding Guarantee}
\begin{theorem}
The protocol ensures mutual binding between $\mathcal{P}$ and $\mathcal{D}$ by requiring both to verify the identity of their counterpart independently.
\end{theorem}
\begin{proof}
The digital twin $\mathcal{D}$ proves knowledge of $sk_d$ through the Schnorr proof, while the physical entity $\mathcal{P}$ exposes $H_1(\mathcal{S}_p)$, which is matched against the registered $pk_p$ via:
\begin{equation}
    pk_p \stackrel{?}{=} g^{H_1(\mathcal{S}_p)}
\end{equation}
This dual verification enforces bidirectional trust, as both entities must possess valid and cryptographically linked credentials. If either check fails, the session is aborted, preventing unauthorised bindings.
\end{proof}

\subsection{Forward Secrecy}
\begin{theorem}
The derived session key $K_{pd} = H_1\big((pk_p \cdot g^{r_p})^{sk_d} \parallel \zeta\big)$ ensures forward secrecy under the CDH/DDH assumptions.
\end{theorem}
\begin{proof}
Even if $sk_d$ is later compromised, the adversary cannot compute $K_{pd}$ for past sessions without knowledge of the ephemeral randomness $r_p$ generated by $\mathcal{P}$ in each session. The term $(pk_p \cdot g^{r_p})^{sk_d}$ combines static and ephemeral secrets. Under the CDH/DDH assumptions, recovering this value without $r_p$ is infeasible, ensuring that $K_{pd}$ remains indistinguishable from random in all previous sessions.
\end{proof}

\subsection{Resistance to KCI and MITM Attacks}
\begin{theorem}
The protocol is resistant to Key Compromise Impersonation (KCI) and Man-in-the-Middle (MITM) attacks, under the assumption that $\mathcal{S}_p$ is non-extractable and $H_1$ is secure.
\end{theorem}
\begin{proof}
In a KCI scenario, an adversary with access to $sk_d$ cannot impersonate $\mathcal{P}$, as the secure identity $\mathcal{S}_p$ is rooted in tamper-resistant hardware and cannot be derived or emulated. For MITM resistance, any modification to the Schnorr transcript $(\alpha, c, z)$ will be detected during verification due to the binding to $pk_d$ and $\zeta$. Any change will lead to failure of Equation~\ref{eq:schnorr_verification}, thus preventing successful interception.
\end{proof}

\subsection{Session Key Security under the ROR Model}
\begin{theorem}
The session key $K_{pd}$ is secure under the ROR model. No PPT adversary can distinguish $K_{pd}$ from random with non-negligible advantage.
\end{theorem}
\begin{proof}
In the ROR game, the adversary $\mathcal{A}$ issues a test-session query and receives either the real session key or a random value. Its goal is to guess the hidden bit $b$ with advantage:
\begin{equation}
    Adv_{\mathcal{A}}^{ROR} = \Big|\Pr[b' = b] - \tfrac{1}{2}\Big|
\end{equation}
Since $(pk_p \cdot g^{r_p})^{sk_d}$ follows the static–ephemeral Diffie–Hellman structure, the DDH assumption guarantees that this value is indistinguishable from random. The hash function $H_1$ further amplifies this property under the Random Oracle Model. Thus, $Adv_{\mathcal{A}}^{ROR}$ is negligible.
\end{proof}

In our system, the role of $\mathcal{CA}$ is limited to initialisation; no runtime interaction is required. If needed, standard approaches such as threshold or federated CAs can mitigate single-point-of-failure risks.

\section{Performance Evaluation}
\label{sec:performance}
To validate efficiency and robustness, we built a simulation environment reflecting smart city edge conditions.
We used a lightweight Python-based simulation platform (Python 3.9.2) running on Ubuntu 20.04 LTS that integrates modular cryptographic libraries (PyCryptodome 3.18.0) \cite{PyCryptodome}, supporting Schnorr zero-knowledge proofs and ECC. Python's native \texttt{threading} module was used to simulate concurrent operations.
Each entity, $\mathcal{P}$ and $\mathcal{D}$, was instantiated on separate execution threads, with synthetic delays injected to reflect network latency and adversarial interference. The underlying cryptographic operations were executed over the secp256r1 (P-256) elliptic curve, with SHA-256 \cite{NISTFIPS186-4} employed for both hash functions $H_1$ and $H_2$. The cyclic group $\mathbb{G}$ was defined with a generator $g$ of prime order $q = 2^{256} - 189$, consistent with ECC standards. Each simulation was executed over 5000 protocol sessions on a baseline hardware configuration comprising an Intel i7 processor (2.4GHz) and 16GB of RAM.

\begin{figure}[t]
    \centering
    \includegraphics[width=3.5in]{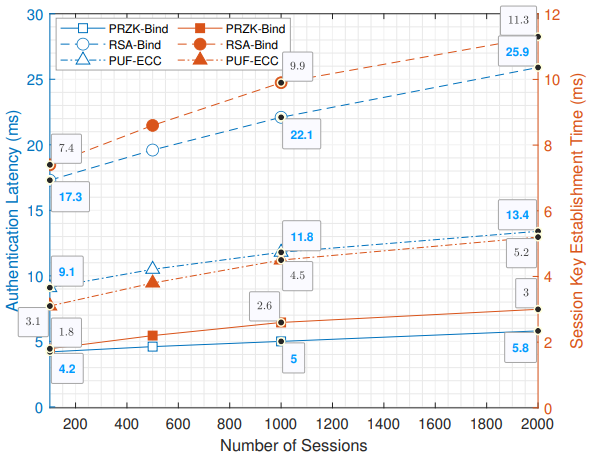}
    \caption{The authentication latency and session key establishment time comparison.}
    \label{fig:auth}
\end{figure}
\begin{figure}[t]
    \centering
    \includegraphics[width=3.5in]{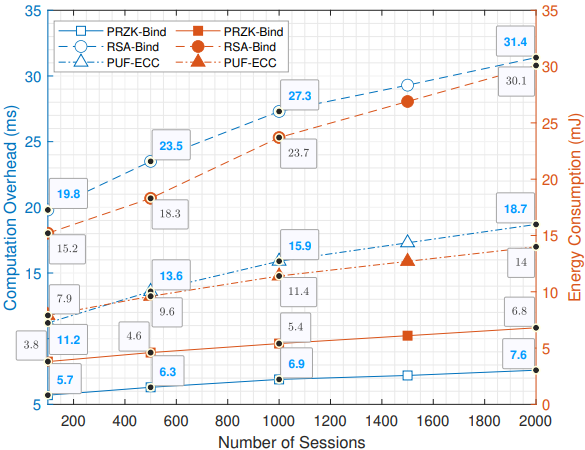}
    \caption{The comparison of computation overhead and energy consumption.}
    \label{fig:overhead}
\end{figure}

We compared the proposed protocol against two baselines widely discussed in the literature: a cryptography-intensive scheme named RSA-Bind, and a lightweight physical identity-based approach named PUF-ECC.
The RSA-Bind protocol, inspired by the design in~\cite{RSAKA25}, integrates key-policy attribute-based encryption (KP-ABE), bilinear pairings, and elliptic curve operations to enable fine-grained access control. However, to reflect practical edge deployment conditions, we implemented the scheme with a five-attribute policy and 160-bit security parameters, following the original specification.
On the other hand, the PUF-ECC protocol, adapted from~\cite{VeDTAut21}, leverages PUFs to derive hardware-tied identities, combined with ECC-based session key establishment. While this design offers low computational cost, it lacks strong guarantees on live identity binding and introduces variability due to PUF response delays.
All protocols were tested under identical and realistic smart city edge conditions to ensure a fair and rigorous evaluation. Each authentication session incorporated synthetic network latency (10–20 ms) and concurrent computational tasks to emulate typical edge workloads. For security assessment, adversarial attempts were systematically injected at a ratio of 10\% across 5000 sessions (4500 legitimate and 500 adversarial). This controlled setup enabled a reliable comparison of performance and security, covering both normal operations and adversarial scenarios.

We first investigated authentication latency and session key establishment performance. As shown in Fig.~\ref{fig:auth}, PRZK-Bind consistently outperforms prior works in both metrics. Compared to RSA-Bind, our protocol achieves up to 4.5 times lower authentication latency and reduces session key establishment time by nearly 3.5 times. Compared to PUF-ECC, PRZK-Bind still provides more than 2 times faster authentication and session key establishment.
This improvement comes from its simple design. PRZK-Bind only uses a single Schnorr step and a quick ECDH computation. By contrast, RSA-Bind is slowed down by complex KP-ABE and pairing operations, while PUF-ECC adds delay due to PUF-based processing.
As shown in Fig.~\ref{fig:overhead}, PRZK-Bind delivers the lowest computation and energy overhead across all protocols, thanks to its streamlined design, which avoids complex cryptographic layers and instead relies on lightweight Schnorr-based proofs combined with elliptic curve operations. Specifically, it achieves up to 4 times lower processing time and energy use than RSA-Bind, which suffers from expensive attribute-based cryptographic operations. Similarly, PRZK-Bind performs approximately 2 times better than PUF-ECC, which incurs additional delays and energy costs due to PUF reading and ECC-based processing.
%
\begin{table}[t]
\centering
\caption{False Acceptance Rate (FAR) Comparison}
\label{tab:far}
\begin{tabular}{c c c c}
    \hline \hline
    \begin{tabular}[c]{@{}c@{}}Adversarial\\Attempts\end{tabular} & PRZK-Bind (\%) & RSA-Bind (\%) & PUF-ECC (\%)\\
    \hline \hline
    100  & 0.00 & 0.11 & 0.07 \\
    500  & 0.01 & 0.18 & 0.13 \\
    1000 & 0.01 & 0.26 & 0.19 \\
    2000 & 0.01 & 0.34 & 0.24 \\
    \hline \hline
\end{tabular}
\end{table}
%
Table~\ref{tab:far} presents the FAR measured under adversarial conditions. PRZK-Bind achieved the best results, maintaining FAR consistently below 0.02\%, even during high attack volumes. This is approximately 10 to 20 times lower compared to RSA-Bind, which reached up to 0.34\%, and about 5 to 10 times lower than PUF-ECC, which peaked at 0.24\%. 
This advantage comes from PRZK-Bind’s strict zero-knowledge authentication, which leaves no room for unauthorised responses. In contrast, RSA-Bind’s attribute checks were vulnerable to borderline cases, while PUF-ECC suffered from unstable identity inputs. 
Overall, PRZK-Bind combines fast authentication, low resource consumption, and strong security, offering a lightweight yet resilient framework for future smart city deployments.

\section{Conclusion}
\label{sec:conclusion}
In this work, we proposed PRZK-Bind, a lightweight and decentralised authentication protocol designed to establish secure and verifiable bindings between physical entities and their digital twins at the smart city edge. Unlike existing solutions, PRZK-Bind ensures live physical-digital correspondence without relying on pre-shared secrets or centralised authorities. Through a Schnorr-based zero-knowledge proof combined with efficient elliptic curve operations, our solution achieves strong security guarantees with minimal latency and overhead. Extensive security analysis and simulation results confirmed its robustness against impersonation and replay attacks, while significantly outperforming current methods in terms of efficiency and reliability. Overall, PRZK-Bind provides a practical and secure foundation for future autonomous and resilient digital twin-enabled smart cities.

\section*{Acknowledgment}
This work was partially supported by The Scientific and Technological Research Council of Turkey (TUBITAK) 1515 Frontier R\&D Laboratories Support Program for BTS Advanced AI Hub: BTS Autonomous Networks and Data Innovation Lab. Project 5239903.

\bibliographystyle{IEEEtran}


\end{document}